\documentclass[acmsmall,screen,preprint]{acmart}
\usepackage{graphicx} 

\newcommand{\toolname}{\textit{PromptPex}}
\newcommand{\numBenchmarks}{twenty-two}
\newcommand{\numTestModels}{four}
\newcommand{\gptthreefive}{\textit{gpt-3.5-turbo}}

\newcommand{\rulemodel}{\textit{gpt-5\_2025-08-07}}
\newcommand{\basemodel}{\textit{gpt-5\_2025-08-07}}
\newcommand{\evalmodel}{\textit{o4-mini\_2025-04-16}}

\newcommand{\llmjudge}{\textit{LLM-as-a-judge}}
\newcommand{\emphbf}[1]{\emph{\textbf{#1}\xspace}}
\newcommand{\mypara}[1]{\smallskip\noindent\emphbf{#1}\xspace}

\newcommand{\testmodels}{\textit{gpt-oss}, \textit{gemma2:9b}, \textit{qwen2.5:3b}, \textit{llama3.2:1b}}
\newcommand{\localmodels}{\textit{gpt-oss}, \textit{gemma2:9b}, \textit{qwen2.5:3b}, \textit{llama3.2:1b}}

\newcommand{\PUT}{\textit{PUT}}
\newcommand{\IS}{\textit{IS}}
\newcommand{\OR}{\textit{OR}}

\usepackage{xcolor}
\usepackage{soul}
\usepackage{xspace}
\usepackage{microtype}

\definecolor{bluebg}{RGB}{230, 240, 255}
\definecolor{bluetext}{RGB}{0, 51, 153}

\definecolor{purplebg}{RGB}{235, 230, 250}
\definecolor{purpletext}{RGB}{102, 0, 153}

\definecolor{redbg}{RGB}{255, 220, 220}
\definecolor{redtext}{RGB}{153, 0, 0}

\definecolor{greenbg}{RGB}{230, 255, 230}
\definecolor{greentext}{RGB}{0, 80, 0}

\definecolor{cyanbg}{RGB}{225, 250, 250}
\definecolor{cyantext}{RGB}{0, 102, 102}

\definecolor{yellowbg}{RGB}{255, 250, 220}
\definecolor{yellowtext}{RGB}{153, 102, 0}

\definecolor{violetbg}{RGB}{240, 230, 255}
\definecolor{violettext}{RGB}{102, 0, 153}

\usepackage{tcolorbox}
\usepackage{caption}
\usepackage{hyperref} 
\usepackage{float}    
\usepackage{multirow}
\usepackage{multicol}
\usepackage{booktabs}
\usepackage{colortbl}
\usepackage{subcaption}
\usepackage{enumitem}

\definecolor{teal}{HTML}{0277BD}    
\definecolor{coral}{HTML}{C2185B}   

\newcommand{\cellcolorpercent}[2]{
    \ifdim #1pt < #2pt
        \cellcolor{coral!15}#1\%
    \else
        \cellcolor{teal!15}#1\%
    \fi
    & 
    \ifdim #2pt < #1pt
        \cellcolor{coral!15}#2\%
    \else
        \cellcolor{teal!15}#2\%
    \fi
}

\usepackage{etoolbox}

\newbool{ignoreContent}

\boolfalse{ignoreContent}

\newenvironment{icse}
 {\ifbool{ignoreContent}
   {\comment}
   {\color{red}}}
 {\ifbool{ignoreContent}
   {\endcomment}
   {}}

\newcommand{\icsein}[1]{%
  \ifbool{ignoreContent}
    {}
    {\textcolor{red}{#1}}%
}

\newtcolorbox{quotefigurebox}{
  colframe=gray, colback=white, arc=5mm,
  boxrule=0.5pt, left=1pt, right=1pt, top=0pt, bottom=0pt,
  sharp corners,
}

\usepackage{listings}
\usepackage[export]{adjustbox}

\title{\toolname{}: Automatic Test Generation for Language Model Prompts}

\author{Reshabh K Sharma}
\authornote{Work done while at Microsoft Research.}
\affiliation{%
  \institution{University of Washington}
  \city{Seattle}
  \state{Washington}
  \country{USA}}
\email{reshabh@cs.washington.edu}

\author{Jonathan de Halleux}
\affiliation{%
  \institution{Microsoft Research}
  \city{Seattle}
  \state{Washington}
  \country{USA}}
\email{jhalleux@microsoft.com}

\author{Shraddha Barke}
\affiliation{%
  \institution{Microsoft Research}
  \city{Seattle}
  \state{Washington}
  \country{USA}}
\email{sbarke@microsoft.com}

\author{Dan Grossman}
\affiliation{%
  \institution{University of Washington}
  \city{Seattle}
  \state{Washington}
  \country{USA}}
\email{djg@cs.washington.edu}

\author{Benjamin Zorn}
\affiliation{%
  \institution{Microsoft Research}
  \city{Seattle}
  \state{Washington}
  \country{USA}}
\email{Ben.Zorn@microsoft.com}

\date{September 2024}

\begin{document}

\begin{abstract}
Large language model prompts differ from traditional code in many ways and require new
approaches to ensure that they are robust.
For example, unlike traditional software the output of a prompt depends on the AI model that interprets it. Also, while natural language prompts are easy to modify, the impact of updates is harder to predict. New approaches to testing, debugging, and modifying prompts with respect to the model running them are required.

To address some of these issues, we developed {\toolname}, an LLM-based tool to automatically generate and evaluate unit tests for a given prompt. {\toolname} extracts input and output specifications from a prompt and uses them to generate diverse, targeted, and valid unit tests. These tests are instrumental in identifying regressions when a prompt is changed and also serve as a tool to understand how prompts are interpreted by different models. We use {\toolname} to generate tests for {\numBenchmarks} benchmark prompts and evaluate the quality of the generated tests by seeing if they can cause each of {\numTestModels} diverse models to produce invalid output.  {\toolname} consistently creates tests that result in more invalid model outputs than a carefully constructed baseline LLM-based test generator. Furthermore, by extracting concrete specifications from the input prompt, {\toolname} allows prompt writers to clearly understand and test specific aspects of their prompts. The source code of {\toolname} is available at \href{https://anonymous.4open.science/r/prompttest-83ED}{https://anonymous.4open.science/r/prompttest-83ED}.
\end{abstract}

\maketitle


\section{Introduction}
Large language models (LLMs) are used in many applications beyond chatbots and prompts for these models are integrated into software applications as code-like artifacts.
These prompts behave much like traditional software in that they take inputs, generate outputs, and perform a specific function~\cite{prompts-are-programs}.
They are also often part of complex chain of control flow that combines LLM-driven prompts and regular code supported by popular frameworks such as langchain~\cite{langchain}.
Such prompts will become an integral part of many code bases in the future because they use the power of AI models and can perform common tasks such as summarization, classification, and evaluation that traditional software is unable to do.  Furthermore, as AI models continue to diversify and become more efficient and effective, software that uses them will benefit. 

\begin{figure}
    \centering         
    \includegraphics[clip, trim=0.1cm 7.1cm 8.87cm 3.10cm, width=\linewidth]{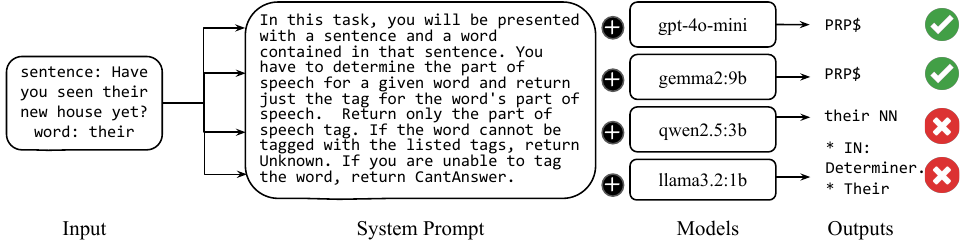}
    \caption{Example illustrating the use of {\toolname{}} to test a given prompt against different models.
    For a given prompt (labeled System Prompt) and {\toolname{}}-generated test input (on the left), 
    the resulting output differs
    depending on what AI model is used to interpret it. {\toolname{}} automatically
    generates test input based on the prompt and evaluates whether the output is compliant with what the prompt specifies. Because the prompt specifies "Return only the part of speech tag" the lower two models produced non-compliant output.}
    \label{fig:multiple_model_output}
\end{figure}

While prompts are becoming a key element of software code bases, they have both similarities and differences from traditional software.  Similarities include taking input, generating output, and performing a transformation much like an ordinary function.  However, there are differences that create major new software engineering challenges. First, the output of a prompt is inherently non-deterministic due to the nature of the underlying AI model and inference engine that interprets it.  Significant effort has been invested in ensuring the model output conforms, at least syntactically, to a given specification~\cite{openai-json}. Also, while prompts are easy for non-programmers to write, the effect of small changes to a prompt are unpredictable, leading to challenges making robust prompt edits.

Second, unlike traditional software, where the behavior of a function depends only on the well-defined specification of the hardware that it is compiled to, the output of a prompt depends on the model that interprets it.  As a result, if the AI model changes, the result of the prompt may change, sometimes dramatically~\cite{retain, prompt-migration-jahani2024generative}. Application developers have strong motivations to change the underlying AI model their application uses because new models are being developed and released that are more efficient, more capable, able to run locally, available as open source, etc. As a result, it is critical that an application developer can quickly understand the implication of changing the underlying model on the behavior of a specific prompt. (Figure~\ref{fig:multiple_model_output} illustrates a concrete example of this).

\mypara{Our Solution: {\toolname{}}}
While many new software engineering practices will need to be adapted to build robust AI software\footnote{We refer to AI software as any software that uses a generative AI model at runtime.}, in this paper we focus on helping developers test and evaluate individual prompts.  Specifically our tool, \toolname{}, takes a prompt as input and automatically generates and evaluates test cases that explore whether a given model interprets the prompt as directed. 

{\toolname{}} uses an LLM to extract explicit specifications from an input prompt (the {\em Prompt Under Test, or PUT}) that capture the user intent in simple and concrete terms.  A key element of our approach is to extract a projection of the prompt as a set of independent, concrete, checkable {\em output rules (OR)} that are then used to create targeted tests. Many prompts used in commercial applications have natural language statements that express such rules.  For example, a prompt might contain  phrases like "Ensure that..." or "The output must ..." that translate directly into our output rules.

From this extracted specification, we use an LLM to generate test cases focused on exploring whether the PUT with a given model ({\em the Model Under Test, or MUT}) adheres to the specification. We then run our test cases with different MUTs to generate model-specific outputs for a given test case, as shown in Figure~\ref{fig:multiple_model_output}.
Because the tests were generated with the specification in mind, we can then automatically evaluate whether a given output from a MUT complies with the specification.  While {\toolname{}} cannot be used to automatically generate test cases that explore the full functionality of a given PUT, it is still valuable in creating test cases that break the requirements of our extracted specification.  

{\toolname{}} is both easy to use and provides immediate insights into potential issues that can arise from the language in the PUT, the MUT, or the combination of the two. To evaluate our approach, we collected a suite of {\numBenchmarks} benchmark PUTs, created tests using both {\toolname{}} and a baseline LLM-based test generator, ran each test with {\numTestModels} MUTs, and evaluated the tests using an LLM to determine if the outputs were compliant with requirements specified in the PUT.  We consider the generation of tests that are non-compliant as more successful because a test that causes a model to generate a non-compliant output indicates a problem that should be addressed.  Our results show that {\toolname{}} consistently generates more non-compliant tests than our baseline test generator and also clearly distinguishes relative model capabilities for the given prompt.

\mypara{Contributions} This paper makes the following contributions.
\begin{enumerate}
\item \textbf{Systematic and Automated Test Generation for AI Model Prompts:}     
    Our work is the first to focus on the specific problem of automated test generation for prompts and creating tests that allow a developer to understand the behavior of their prompt across multiple models. Prior work has focused on prompt optimization for a model (e.g., ~\cite{prompt-opt-pryzant2023automatic}) and generating unit test cases for programs ~\cite{molina2024test, xue2024llm4fin, zhang2024exploring, bhatia2024unit}.
    
    \item \textbf{Specification Extraction from Prompts for Testing: }
    To generate effective tests, we define a new approach to extract an input specification and output rules that capture targeted properties of prompts.  These generated artifacts can be used both to automatically generate tests and help the prompt developer test, refine, and migrate their prompts to new models.

    \item \textbf{Evaluation of Test Generation and Model Compliance with Specifications:} We measure the effectiveness of {\toolname} using a benchmark suite of {\numBenchmarks} benchmark prompts running the generated tests on {\numTestModels} diverse AI models.  We compare with a sophisticated baseline LLM-based test generator and show that our approach extracting a specification from the input prompt results in tests that are more likely to cause non-compliant outputs across all the models tested.
\end{enumerate}
\section{Motivating Example}

Chatbots, for example for customer support, are a major application of LLMs where natural language prompts describe the bot's behavior.
These descriptions, or meta/system prompts, specify the role and other properties of the chatbot.
They limit the domain of input and restrict the possible outputs that can be generated.
Using prompts to configure a chatbot makes customization and modification easy, contributing to their widespread success.
These prompts are designed to directly interact with user input, engage in conversations, and provide information to users.

The prompts we focus on, which are used as software artifacts, share some similarities but also have distinct features from chatbot prompts.
First, prompts used in codebases act like programs, often with well-defined inputs and outputs.
This may include both syntactic and semantic restrictions.
While LLMs are generally good at handling inputs that do not adhere to format, the output must still be well-formed for other parts of the codebase, which may be traditional, to handle it effectively.
These prompts exhibit constructs similar to programming constructs, such as performing an early return for particular input, similar to an if-then-return in traditional programming.
They can also include features like complex control flow, multiple returns, assertions, and constraints.
All these features are described in natural language and can be easily modified.

Unlike in chatbot applications, where users can often provide any input and the chatbot needs to handle it, the input domain of a prompt embedded inside a codebase is more narrowly defined.
The prompts embedded in software expect well-defined input. The constraint over the input can be explicitly stated within the prompt itself to filter out invalid input or be implicit in the codebase so that invalid input is never sent to the prompt. 
These prompts do not operate in isolation. They are part of complex logic pipelines where the output from a prompt or a traditional program can be fed into one another. These pipelines can be thoroughly layered, made up of multiple prompts and programs processing data to create a single response.
The input to a prompt may be preprocessed to optimize efficiency or to filter out invalid data. Likewise, the output can be validated, and if necessary, the prompt can be executed again with the same input to generate a well-formed response that can be correctly parsed by dependent components~\cite{TypeChat, pydanticValidatorsPydantic}.

The prompts that are part of codebases are rich in program-like constructs. We used a collection of such prompts in our evaluation. We use a simpler prompt as our running example as shown in Figure~\ref{fig:pos_prompt}. Although this prompt is simple, it includes program-like constructs similar to those found in complex prompts and is representative in highlighting the challenges faced by the prompts being used in traditional software pipelines.

\subsection{Prompt Under Test (\PUT{})}
The prompt in Figure \ref{fig:pos_prompt} is the part-of-speech classification (POS) prompt\footnote{A modified version of the prompt from~\cite{prompt-opt-sammo-schnabel2024symbolic}.}, which classifies a word into a POS tag. We have truncated the list of speech tags with their description for brevity.

\begin{figure}[ht]
\centering 
\small
\begin{quotefigurebox}
In this task, you will be presented with 
\textcolor{purpletext}{\sethlcolor{purplebg}\hl{a sentence and a word contained in that sentence.}}
You have to 
\textcolor{greentext}{\sethlcolor{greenbg}\hl{determine the part of speech for a given word}} 
and return just the tag for the word's part of speech.
\textcolor{redtext}{\sethlcolor{redbg}\hl{}}
Return only the part of speech tag.
\textcolor{cyantext}{\sethlcolor{cyanbg}\hl{If the word}}
cannot be tagged with the listed tags, 
\textcolor{yellowtext}{\sethlcolor{yellowbg}\hl{ return}}
\textcolor{redtext}{\sethlcolor{redbg}\hl{Unknown. }}
\textcolor{cyantext}{\sethlcolor{cyanbg}\hl{If you are}}
unable to tag the word,
\textcolor{yellowtext}{\sethlcolor{yellowbg}\hl{ return}}
\textcolor{redtext}{\sethlcolor{redbg}\hl{ CantAnswer.}}
Here is the alphabetical 
\textcolor{violettext}{\sethlcolor{violetbg}\hl{list of part-of-speech tags}}
used in this task:
CC: Coordinating conjunction, 
CD: Cardinal number, 
...
\end{quotefigurebox}
\caption{Part-of-Speech Prompt}
\label{fig:pos_prompt}
\end{figure}

The POS prompt has the following program-like constructs:
\begin{itemize}
    \item \textcolor{purpletext}{\sethlcolor{purplebg}\hl{\textbf{Input:}}} It takes a sentence and a word as input, specifying that the word must be present in the sentence.
    \item \textcolor{redtext}{\sethlcolor{redbg}\hl{\textbf{Output:}}} It defines output as POS tag, \texttt{Unknown}, or \texttt{CantAnswer}.
    \item \textcolor{greentext}{\sethlcolor{greenbg}\hl{\textbf{Computation:}}} It describes how the output should be computed, which in this case is simply tagging the word as a POS tag.
    \item \textcolor{cyantext}{\sethlcolor{cyanbg}\hl{\textbf{Control flow:}}} There are multiple if-then constructs in the prompt, similar to those found in code.
    \item \textcolor{yellowtext}{\sethlcolor{yellowbg}\hl{\textbf{Early return:}}} Like in programs where the code terminates early, returning an error code, prompts also have multiple early returns to handle corner cases.
    \item \textcolor{violettext}{\sethlcolor{violetbg}\hl{\textbf{Assertions and constraints:}}} Like in programs, assertions and constraints can be defined directly in the prompt. However, they need not be explicitly implemented. For example, the tag must be from the list of tags provided.
\end{itemize}

These prompts function like programs, featuring well-defined input and output specifications. For instance, when given the input \textit{quick brown fox jumps over the lazy dog; quick}, the output generated is \textit{JJ}. This input might originate from another prompt or program, and similarly, the output can be passed on to another component.
While the POS prompt may align with what the prompt developer intended, ensuring that the LLM interpreting the prompt can precisely understand the intent remains a challenge. Developing these prompts is demanding because they require more rigorous testing than traditional software due to the vast potential input domain, even with constraints in place. Moreover, the model's behavior can vary unpredictably based on the input, which can be neither fully understandable nor predictable. 
LLMs often struggle to consistently follow provided instructions, but these instructions can be made more precise to accurately match the intended purpose, thereby minimizing unintended consequences. As models improve their ability to follow instructions, the alignment between the prompt developer's intent and the model's actual execution behavior will become increasingly critical.

As an example of the challenges in writing an effective prompt, the POS prompt above sometimes generates a POS tag along with a description of the reasoning steps taken to arrive at that decision. This issue sometimes occurs even with {\it gpt-oss}, our most effective model-under-test. We observed this behavior due to ambiguity in what is allowed as output. To understand the ambiguity, note that the prompt requires a POS tag (e.g., NN) and not the word describing the part of speech (e.g., noun).  Specifically, the prompt says "Return only the part of speech tag." The ambiguity arises because sometimes a model interprets this rule as applying only to tag (interpreting the prompt as "Return only the part of speech tag and not the word describing the part of speech")
and does not forbid adding an explanation about the reasoning behind the choice. This demonstrates the intricacies of prompt development and the value of both being more explicit about what is expected and testing model behavior rigorously. 

The same prompt on {\it gpt-oss} most of the times only outputs the tag, but on {\it gpt-3.5} it often prefixes the tag with \textit{Output:}, illustrating the problem of model portability. A prompt that works correctly on one model may not function as expected on another, highlighting the importance of extensive testing across multiple models.
These challenges underscore the necessity for thorough testing of prompts on various models to gain insights into how prompts behave during execution on different models. This understanding can help prompt developers address problems and optimize their prompts across different platforms.

We developed \toolname{}, a tool designed to help prompt developers better understand the execution behavior of their prompts. It achieves this by first extracting input and output specifications for the prompt under test (\PUT{}), which are assertion-like constraints over the input and output, equivalent to pre- and post-conditions in a program. 
\toolname{} utilizes {\it gpt-oss} to extract these specifications from the prompt. The prompt developer can examine these extracted specifications to compare their understanding with what a SOTA model perceives as expected input and output constraints. 

\subsection{Input Specification (\IS{})}
For the POS prompt, Figure \ref{fig:input_spec_pos_prompt} shows the extracted input specification (\IS{}) by {\it gpt-oss}. 
For the \PUT{}, \IS{} specifies that the input must be a sentence and a word, with constraints like the word being a single word from the sentence. 
Though not explicitly stated in \PUT{}, the model must decide whether to allow compound words (e.g., ice cream). 
The \IS{} highlights this under-specification in the \PUT{}, indicating that the model has assumed it is expecting single words. Compound words might be considered valid input based on the developer, but they would result in undefined behavior for the model and are likely to lead to the output \texttt{Unknown} or \texttt{CantAnswer}.
\begin{figure}[ht]
\centering 
\small
\begin{quotefigurebox}
$\circ$ The input consists of a sentence combined with a specific word from that sentence.

$\circ$ The sentence must contain natural language text.

$\circ$ The word must be a single word from the provided sentence.
\end{quotefigurebox}
\caption{Extracted Input Specification for POS Prompt}
\label{fig:input_spec_pos_prompt}
\end{figure}
\subsection{Output Specification or Rules (\OR{})}
The output specification, or the rules governing the output (\OR{}), captures the output constraints generated for \PUT{}. Figure \ref{fig:output_spec_pos_prompt} shows the extracted (\OR{}) for the POS prompt.
The prompt developer can compare these rules with their understanding. For instance, the listed rules specify that the output must consist solely of the tag, without any additional text or formatting. This can be interpreted to mean that extraneous text, such as descriptions of the tag or formatting details, is not allowed—such as the use of \textit{Output:} seen with \gptthreefive{} but it does not restrict description of the reasoning behind the tag. These rules help the prompt developer understand the potential behavior of the prompt during execution and can be used to tune the \PUT{} to generate specifications that more accurately match the intentions of the prompt developer.

\begin{figure}[ht]
\centering 
\small
\begin{quotefigurebox}
$\circ$ The output must return only the part of speech tag without any additional text or formatting. 

$\circ$ If the given word can be identified with one of the listed part of speech tags, the output must include only the specific tag for that word from the provided alphabetical list.

$\circ$ If the given word cannot be tagged with any of the listed part of speech tags, the output should be the word "Unknown".

$\circ$ If tagging the given word is not possible for any reason, the output should be the word "CantAnswer".
\end{quotefigurebox}
\caption{Extracted Output Rules for POS Prompt}
\label{fig:output_spec_pos_prompt}
\end{figure}

After first generating the \IS{} and \OR{}, 
\toolname{} can then generate unit tests for the \PUT{}. The prompt developer also has the option to edit the extracted specification to add implicit rules that are part of the pipeline in which the prompt is embedded but are not needed within the prompt itself, as the input is preprocessed. For example, the prompt developer can extend the input specification to provide the format for the input, such as \textit{"sentence;word"}.

\subsection{Test Generation}
So far, the prompt developer is able to align the intent with the input and output specifications extracted by \toolname{} for the SOTA model. However, this does not account for how the prompt will actually perform on various inputs, especially when tested on other models. The test inputs generated by \toolname{} gives the prompt developer a test suite designed to cover all the output constraints in the prompt and also to challenge the model to violate the constraints set within the prompt. By running these tests, the prompt developer can identify additional failures. Figure \ref{fig:test_gen_pos_prompt} shows a few tests generated by \toolname{} for the POS prompt. This test suite also serves as a regression test suite for any future modifications to the prompt. With passing tests, the suite now encodes the developer's intentions as tests which were initially present only as specifications.

\begin{figure}[ht]
\centering 
\small
\begin{quotefigurebox}
$\circ$ An aura of mystery surrounded them; aura 

$\circ$ The researchers documented carefully; carefully

$\circ$ This is such a unique perspective; such
\end{quotefigurebox}
\caption{Tests Generated for Part-of-Speech Prompt}
\label{fig:test_gen_pos_prompt}
\end{figure}

These tests can now be executed across different models to compare and analyze the execution behavior of the prompt on other platforms. \toolname{} assists the prompt developer in better understanding the behavior of the prompt by explicitly aligning the developer's intentions with the extracted specification and implicitly through the generated test suite. 

\subsection{Test Evaluation}
\toolname{} supports an automated approach to test evaluation which checks test output not for {\it correctness} but for {\it compliance with the prompt} using an LLM.
One use case for \toolname{} is to generate tests and then allow the user to add those tests to their existing test suite.  Note that \toolname{} does not generate the correct output for each test case, so this scenario requires the user to add the correct output for the generated tests as an additional step.
%
\begin{figure}[ht]
  \centering
  \small
  \includegraphics[clip, trim=0cm 0.40cm 0cm 2cm, width=\textwidth]{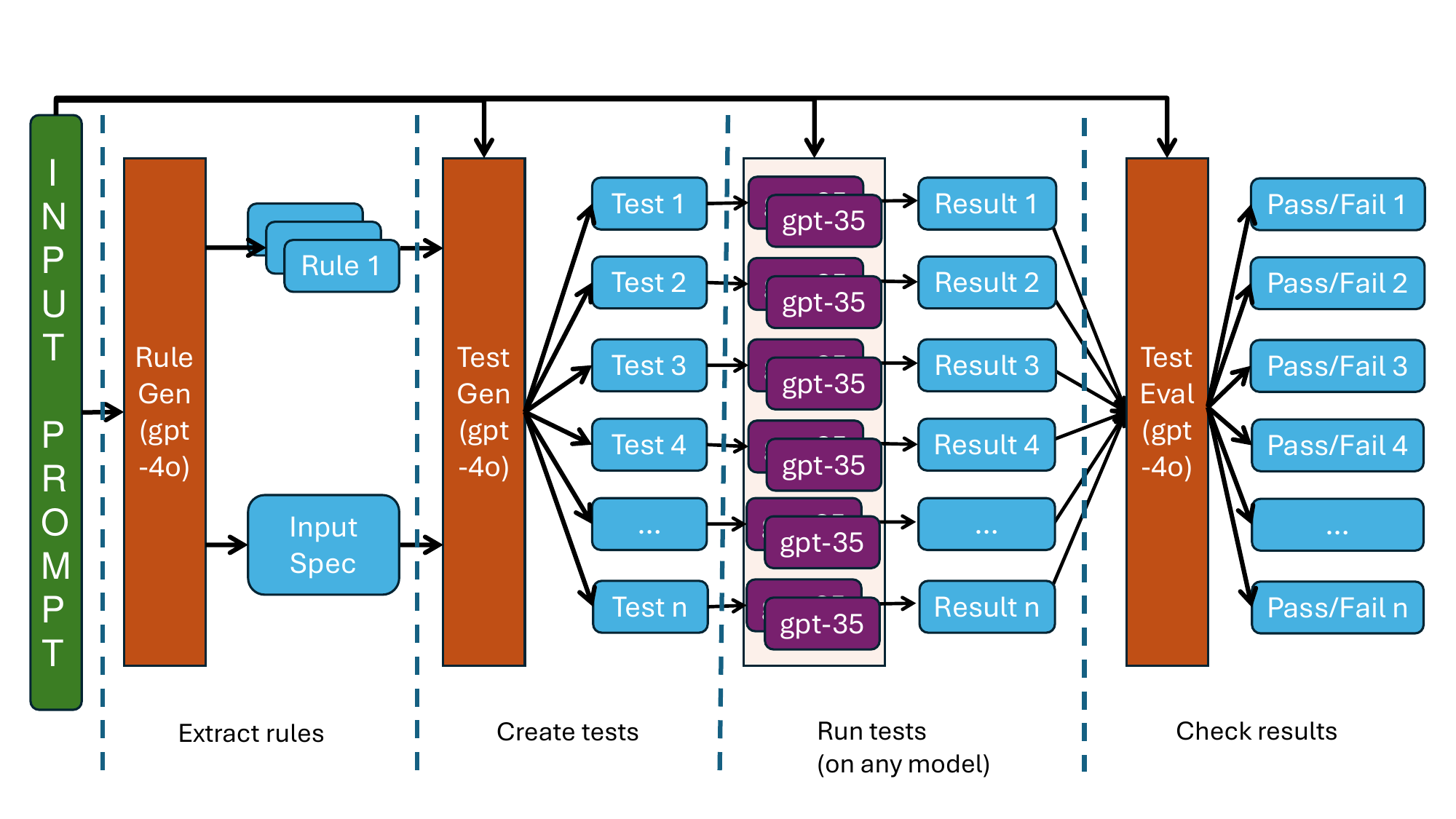}
  \caption{The end-to-end \toolname{} pipeline which
  is implemented using a series of LLM prompts (shown in brown). The user provides their prompt (in green on the left), and \toolname{} automatically generates input and output specifications, and tests (in blue), and can run the generated tests on multiple models (shown in purple). 
  }
  \label{fig:pipeline}
\end{figure}

\section{\toolname{} Design}
\label{sec:design}

In Figure \ref{fig:pipeline}, we present the end-to-end pipeline of \toolname{}. It helps the prompt developer to explore prompts and understand their behavior on different models. 
Below, we discuss each part of the pipeline in detail.

\subsection{Input Specification}
The \IS{} describes the input and the constraints expected by the \PUT{}. The \IS{} defines what constitutes valid input. For the POS prompt, the \IS{}, shown in Figure \ref{fig:input_spec_pos_prompt}, describes the input as a sentence and a word, with the constraint that the word must be present in the sentence. Some prompts may also accept input in the form of a file; in such cases, the \IS{} treats file input as string content and describes the content itself as the input rather than the file.
\toolname{} generates the \IS{} using \rulemodel{}. We frame it as a task to extract the \IS{} to create valid inputs. We restrict any details about the output or how the input will be used in the computation inside the \PUT{}.  In creating the \IS{}, we first extract a description of what the inputs are. If they are composed of multiple components, those components should be listed, and their constraints and properties should be described.
Sometimes the \PUT{} will attempt to handle corner cases. For example, in the POS prompt, if the word cannot be tagged with the listed tags, it returns \texttt{Unknown}. In this case, the \IS{} extractor might incorrectly assume that words which cannot be tagged are not valid inputs. While this may be true, the prompt does not imply that. We explicitly added this case in the \IS{} extractor prompt to consider such inputs as part of the input domain, even if there is a rule against them.
Both \OR{} and \IS{} can be edited by the prompt developer, allowing them to modify the extracted specifications or augment them with constraints from the environment. 

\subsection{Output Specification or Rules}

OR describes the output constraints generated for \PUT{}. Rules are concrete, checkable, general, input-agnostic, and independent constraints over the output
described in natural language and derived from the  \PUT{}.
They are similar to assertions and post-conditions in a program, detailing what the output must be, without regard for how it is generated or derived.
In our POS prompt, the \OR{}, as shown in Figure ~\ref{fig:output_spec_pos_prompt}, includes rules stating that the output should either be a POS tag, \texttt{CantAnswer}, or \texttt{Unknown}. It is important to note that \OR{} do not capture how the output is generated; instead, it focuses solely on constraints over the generated output.
This quality of \OR{} helps in keeping it separate from the input, allowing evaluation of the output based solely on the \OR{}, regardless of input.
The groundedness of the rules within the \OR{} in the prompt is a necessary condition for the \OR{} itself as it implies that all rules are valid and are present in the prompt, while the exhaustiveness in accurately covering all rules from the prompt is the sufficient condition for the \OR{} as no more rules are required.

We use \rulemodel{} for extraction of the \OR{} from \PUT{}. We frame it as a task to extract the rules for output validation such that the inputs are not available. We enforce the following properties during the extraction of the rules: 1) If an example is present in the prompt, do not generate rules specifically for that example. Generalize them so that they will applicable for other possible inputs, 2) Rules must be clear, concrete and independent from each other such that they can individually be used to validate the output. and 3) They should not contain any information about how the output depends on the input or how the output is computed. \footnote{We limit our scope to output compliance testing where we only validate the output unlike functional testing where we validate the output for the given input.}

\subsection{Test Generation}

Tests for \PUT{} are generated by \toolname{} using \OR{} and \IS{}. The \OR{} is utilized to create directed tests that challenge the model to adhere correctly to each rule, while the \IS{} is used to create valid test cases. The input domain specified by \IS{} is usually extensive, as these inputs are in natural language, which can be presented in multiple forms. It is also non-trivial to determine which part of the prompt a given input covers, leaving us without a notion of coverage in prompts unlike what we have in programs.

\subsubsection{Exhaustiveness}
Given the vast range of possible inputs, creating an exhaustive set of test cases is challenging. To achieve this, we generate tests for each rule in \OR{}. We argue that if our \OR{} is exhaustive (completely covers the prompt), the tests generated for the rules in the \OR{} are also exhaustive.
\toolname{} not only generates tests but also associates each test with a specific rule, providing reasoning for its creation. Beyond exhaustiveness, this approach allows for various analyses of the tests, as they are directly linked to a rule. We use this approach to attempt to develop an exhaustive test suite.  Other possible future uses include modular updates to the test suites, allowing the addition of new tests for new rules and the removal of old tests linked to obsolete rules.

\subsubsection{Generating challenging tests}
\label{ssec:generate-challenging-tests}

Since we generate tests per rule, we can create tests that explicitly challenge these rules. We accomplish this while ensuring our test generator remains unaware of any properties of the rule itself; for a given prompt, our test generator will produce a valid test for any rule.  

\begin{figure}[ht]
    \centering     \includegraphics[clip, trim=0cm 9cm 9cm 0cm, width=\linewidth]{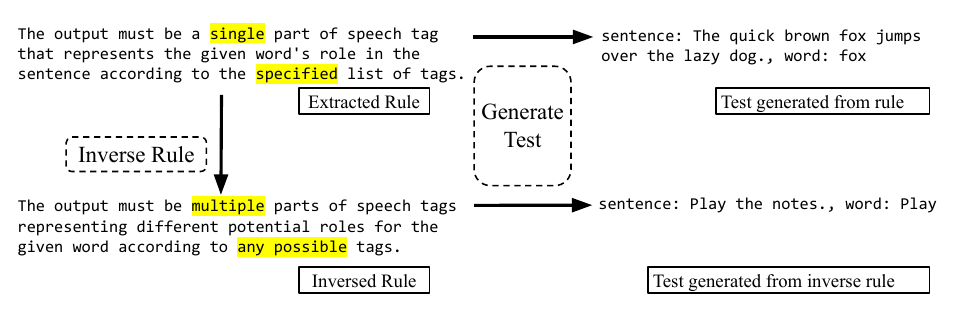}
    \caption{Example illustrates
 inverse rules for test generation. Because inverse rule
    was used to generate the test in the lower half of the figure, the test intentionally focuses on the word "Play" which can be labeled with multiple parts of speech.}
    \label{fig:inverse_rule_to_tests}
\end{figure}

\textbf{Inverse Rules:} We generate inverse rules from the given rules. The inverse of a rule is a semantic inversion that violates the rule by describing its opposite. For example, if a rule in \OR{} states that the output must always be a tag, it can be inverted into a rule that enforces the output to be the tag along with the actual name of the part of speech. 
\toolname{} uses \rulemodel{} for generating inverses of the given rules. We ask it to generate inverse rules such that they contradict the given rules.
We generate tests for both the rules from the \OR{} and their inverses as shown in Figure \ref{fig:inverse_rule_to_tests}. This approach helps us generate tests that cover intriguing test cases which might cause a given model to violate the \OR{}.

\subsubsection{Valid Tests}

A test may not follow the \IS{} and might create output that violates the rules described in the \OR{} and \PUT{}. Such test cases are not the most valuable, as the prompt developer knows that some of those inputs can be pre-filtered before they are passed to the prompt. However, these test cases might be useful for testing any input validation pipeline that processes the input or for identifying gaps in the assumed input domain. The most valuable tests follow the \IS{} as they are actionable and require fixing, so we use the extracted \IS{} to guide \toolname{} in generating them.
Even in scenarios where the prompt-based system is exposed to raw inputs, the same \IS{} can validate inputs and disregard any input that does not meet the input specification.

The test generator in \toolname{} uses \rulemodel{} and takes a rule (which can also be an inverse rule), \IS{} and \PUT{} as input. It generates a test along with the reasoning of why the model thinks this particular test during execution will comply with the rule it was generated for. We ask the test generator to start with first understanding: What is an input? What are the different components of a valid input? What are the syntax and semantics related constraints from \IS{}? In the prompt, we direct the model to consider these issues during test generation. Each test must be generated such that the expected output or behavior demonstrates adherence to the given rules for a range of scenarios.

\subsection{Test Evaluation: Compliance vs Correctness}
\label{ssect:impl:eval}
The generated test cases can be executed on multiple models, helping developers analyze prompt behavior across diverse inputs.
As mentioned, because we lack the correct output for each test input, our automated strategy for test evaluation is an LLM-based determination of test compliance.  
Compliance is determined by our \llmjudge{}~\cite{llm-as-a-judge} which,  given the original prompt and the test output, determines whether the output complies with the requirements described in the prompt.
As a result, our evaluation explores only a partial understanding of the model output and must be augmented with additional testing.
However, as demonstrated in traditional software engineering, testing the assertions over the output can also be very valuable~\cite{assertions-swe-burnett2003end, assertions-swe-clarke2006historical, assertions-swe-taromirad2024literature}. After the tests are run, the results of our validator can assist the prompt developer in updating the prompt to correctly handle tests with invalid results.

To implement \llmjudge{}, 
we use \evalmodel{} and provide it the output generated by executing a test and the PUT. To allow a fair comparison with a test generator that uses \toolname{}, we ensured that no artifacts specific to \toolname{} are used in the validation and that the test output validator is generic, applicable to any prompt and test output regardless of the input, the method used to generate the test, or the model used to run the test.

This validator is used to evaluate the outputs of tests generated by both the baseline and \toolname{}. We enforce the following properties for this output validator: 1) Keep the evaluation independent of the input, as it will not be provided. 2) Must not speculate, infer, or make any assumptions during the evaluation. 3) Always check for compliance, not correctness.
The validator generates a boolean result (compliant/non-compliant) and explanation for the decision.

\section{Evaluation}
\label{sec:evaluation}

In this section, we evaluate how effective \toolname{} is at generating tests that provide actionable feedback to the developers for improving the prompts. 
Through our experiments, we answer the following research questions:
\begin{enumerate}[label=(\bfseries Q\arabic*)]
\item Do specifications generate better tests?
\item How useful are rules for generating better tests? 
\item Does an input specification help generate valid tests? 
\item Can automatically generated tests help determine if a model is suitable for a prompt?
\end{enumerate}
\subsection{Evaluation Setup}

\subsubsection{Baseline}
In our evaluation, we used a zero-shot LLM-based test generator as the baseline. We use \basemodel{} to generate tests for both the baseline and \toolname{}. We instruct the model to develop multiple test cases by inferring the functional specification and input for the given prompt. We ask it to begin with understanding possible inputs and its components, what the syntax and semantics related constraints are, and possible input scenarios. We enforce the following properties while generating the baseline tests: 
1) The test cases must be designed to validate whether the output properly adheres to description, 2) A good test must always be a valid input meeting the requirements mentioned in the description, 3) The test cases must be diverse and distinct, 4) Each test case must be crafted to rigorously assess whether the output meets the stipulated behavior based on the provided prompt, 5) The input scenarios used for creating the tests must be valid, realistic, and fully comply with the given description, 6) Generate test cases to broadly cover a range of scenarios, including boundary cases, typical cases, and edge cases, to thoroughly evaluate the software's adherence to the description, 7) Each test case should adhere to principles of good software testing practices, emphasizing coverage, specificity, and independence, 8) Focus on creating diverse test cases that effectively challenge the prompt's capabilities by critically assessing potential weaknesses in the handling of inputs by the prompt.

We chose this as our baseline because LLMs are capable of interpreting the prompt and generating test cases for them. LLMs are already useful in generating unit tests for traditional software~\cite{llm-old-swe-test-gen-og, llm-old-swe-test-gen-siddiq2024using, llm-old-swe-test-gen-tang2024chatgpt, llm-old-swe-test-gen-xie2023chatunitest, llm-old-swe-test-gen-yuan2023no, llm-old-swe-test-gen-yuan2024evaluating, llm-old-swe-test-gen-yang2024evaluation}. We ensured that the prompt used to generate the baseline tests is robust and underwent multiple revisions to enhance its ability to generate effective tests. To provide a fair comparison with \toolname{}, we explicitly covered all requirements that are implicitly enforced by \toolname{}, such as generating valid, challenging, and diverse test cases that comprehensively cover the prompt and help identify any flaws. Although there will always be opportunities to improve the baseline prompt, the same holds for the prompts used in \toolname{}. We have refined both approaches sufficiently so that they can be effectively compared.

\subsubsection{Metrics}
We evaluate \toolname{} and the baseline based on their ability to generate more effective tests.

\mypara{\% non-compliance}
We define effective tests as those that expose limitations in the prompt, which means that they result in more failures. We consider non-compliance with the prompt as the metric for test quality. This includes any violations of the rules and constraints described in the prompt. 
This approach is similar to checking all the assertions over the output generated by a traditional program.

We used \llmjudge{} to evaluate whether the output generated by the prompt violates any rules or constraints within the prompt as described in Section~\ref{ssect:impl:eval}. 
We could have also developed a validator that does not use the \PUT{} but instead relies on the \OR{}, as they precisely represent the constraints over the output. However, since the \OR{} is part of \toolname{} and was not provided to the baseline test generator, we refrained from using it to maintain a fair and equal comparison. 
We run the tests generated by \toolname{} and the baseline on the following models: \testmodels{}. We compare the non-compliance of the output from the generated tests. In our evaluation, we consider more test non-compliance as better since it indicates that the tests are more effective at identifying gaps in the prompt’s constraints.

\mypara{Test Validity}
Valid tests follow the input spec (\IS{}) defined by the prompt-under-test (PUT). We use test validity as a metric for \toolname{}'s test quality. Valid tests are important as they represent likely inputs as compared with invalid inputs that may result in non-compliance but will also may be less likely to occur in actual use. We use an \llmjudge{} (\evalmodel{}) comparing the test against the \IS{} to determine if the generated test is valid. 

\mypara{Groundedness and Spec Agreement}
We also evaluate the generated \OR{} as these are directly used to generate tests by \toolname{} and have impact on the quality of the generated tests. 
A rule in the \OR{} is considered grounded if it is present in the \PUT{}. We do not want to generate tests for the rules which are not present in the \PUT{}. We consider a higher groundedness of the rules as an indirect metric for test quality. To determine if the generated rules are grounded in the original \PUT{}, we used \llmjudge{} and ask it to confirm rule groundedness.

\begin{figure}
    \centering
    \footnotesize
    \includegraphics[clip, trim=6.4cm 7.4cm 11.6cm 3.8cm, width=0.7\linewidth]{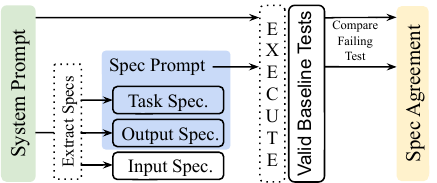}
    \caption{Spec agreement estimation for \toolname{} tests. The specifications are extracted from the prompt and a spec prompt is created by appending the output specification to the task specification. Spec agreement is considered high when the spec prompt and original prompt behave similarly.}
    \label{fig:spec-agreement}
\end{figure}

Another goal for generating \OR{} is to ensure that all important constraints mentioned in the \PUT{} are captured in the \OR{}. We refer to this as spec agreement.
To compute this, we extract a description of how the output must be computed from the \PUT{}, the {\it task specification}. We append the task specification with our extracted \OR{} to derive a \emph{spec prompt} which in theory should capture all the same constraints as the \PUT{}.
We do not use the \IS{} in the spec prompt because \toolname{} generates tests for each rule in the \OR{}. 
To estimate how closely the extracted specification represents the original \PUT{}, we compare the spec prompt  with the original \PUT{}. 
We feed tests generated from our baseline test generator to both the spec prompt and the \PUT{} to compare their behaviors as shown in Figure \ref{fig:spec-agreement}.
We use the baseline tests instead of the \toolname{} tests so that they are not influenced by our process of generating them from the \OR{}.  
The spec prompt and the original \PUT{} must have similar behavior for the generated \OR{} to have high spec agreement. We use cosine similarity of the non-compliance percentage of spec prompt and the original \PUT{} to derive a spec agreement score.

\subsubsection{Benchmarks}
We selected a diverse set of prompts from publicly available sources, focusing on those that are within the scope of \toolname{}. Currently, we support prompts that can accept only a single input. Although this single input can be interpreted by the model to be made up of multiple components, this differentiation is not explicitly present, for example, a single string as input representing a sentence and a word for the speech tag prompt. This limitation makes it unsuitable for prompts requiring multiple embedded inputs. We also only support prompts where the output is independent of the previous outputs, making prompts describing multi-turn conversations for tasks out of scope for \toolname{}. 

\subsubsection{Evaluation Procedure}
We accessed \rulemodel{} and \evalmodel{} through APIs. We kept the temperature 1.0 across all the requests. We used Ollama~\cite{ollama} for the local models, \localmodels{}. We ran all the experiments and hosted the local models on a virtual machine hosted in the cloud. Our system runs on an AMD EPYC 7V13 processor with 220 GB of RAM, 1 TB of SSD, and 4 NVIDIA A100 GPUs with a total of 80 GB of dedicated memory. The VM is running Ubuntu 24.04.3 LTS. We ran each test once per prompt per model.

\subsection{Evaluation Results}
\mypara{Q1: Do specifications generate better tests?}

Our hypothesis is that explicitly extracting output rules and using them to generate tests is more effective at generating tests that will cause non-compliance.  But does our approach provide any benefits over baseline?
To answer this question, we compare the non-compliance of tests generated by \toolname{} and our baseline using 4 different models-under-test and averaged the scores across all the benchmarks. Figure~\ref{fig:prompttest-baseline-average-test-compliance} shows the average non-compliance across the benchmarks for \toolname{} and baseline for each of the different models.  More capable models will have lower non-compliance scores, as we see with the {\it gpt-oss} results when comparing against {\it llama3:2.1b}. Comparing against baseline, we see that the \toolname{} tests have higher non-compliance than baseline tests for every model-under-test, but is particularly more effective for {\it gpt-oss}, where the baseline model has a harder time generating tests that result in non-compliant output.

\begin{icse}
\begin{figure}[ht]
    \centering
    \begin{minipage}{0.48\textwidth}
        \centering
\includegraphics[width=1.0\linewidth]{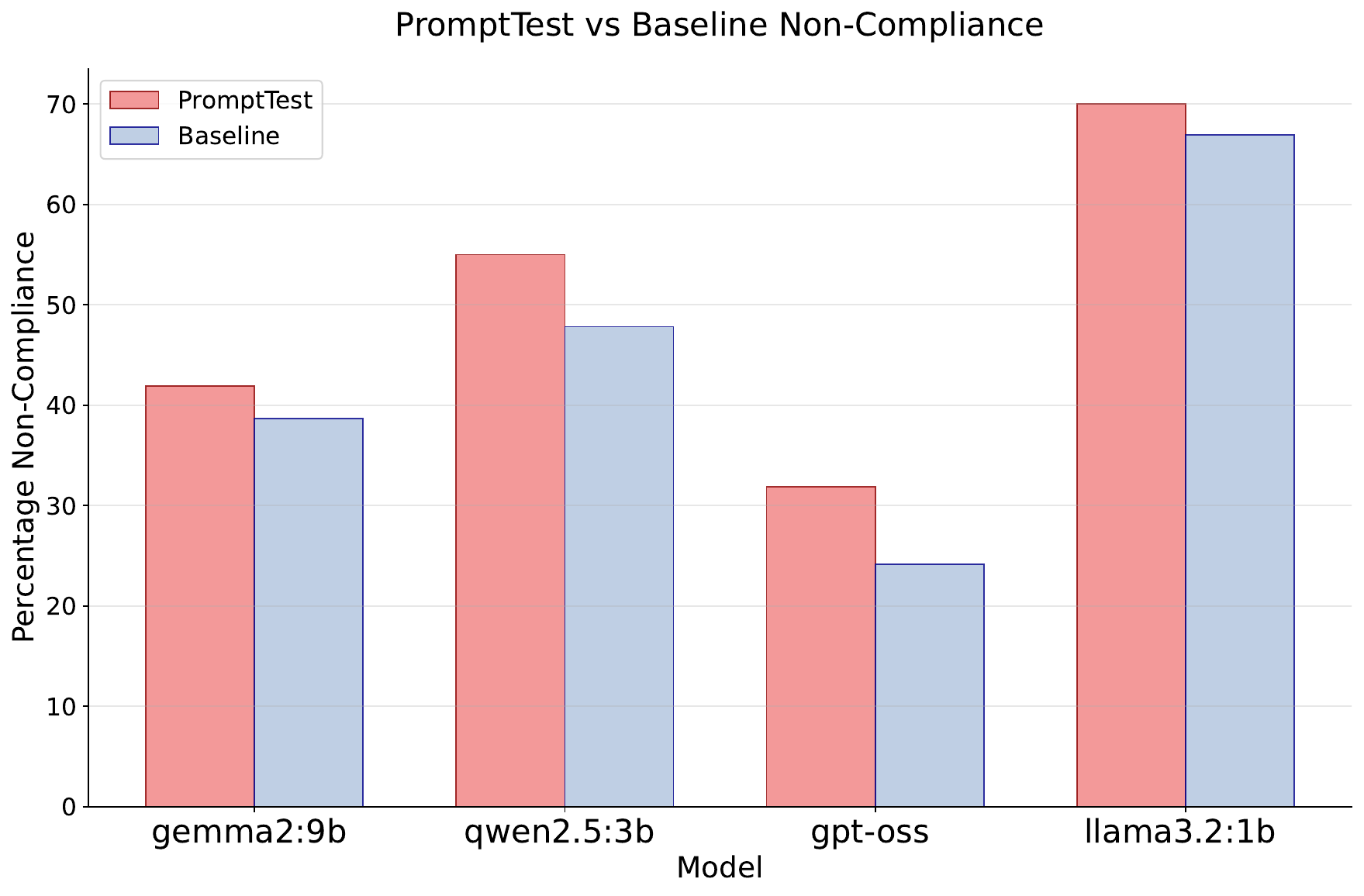}
        \caption{Average \% Test Non-Compliance of tests generated by \toolname{} and Baseline. Higher is better.}
        \label{fig:prompttest-baseline-average-test-compliance}
    \end{minipage}
    \hfill
    \begin{minipage}{0.48\textwidth}
        \centering
        \includegraphics[width=1.0\linewidth]{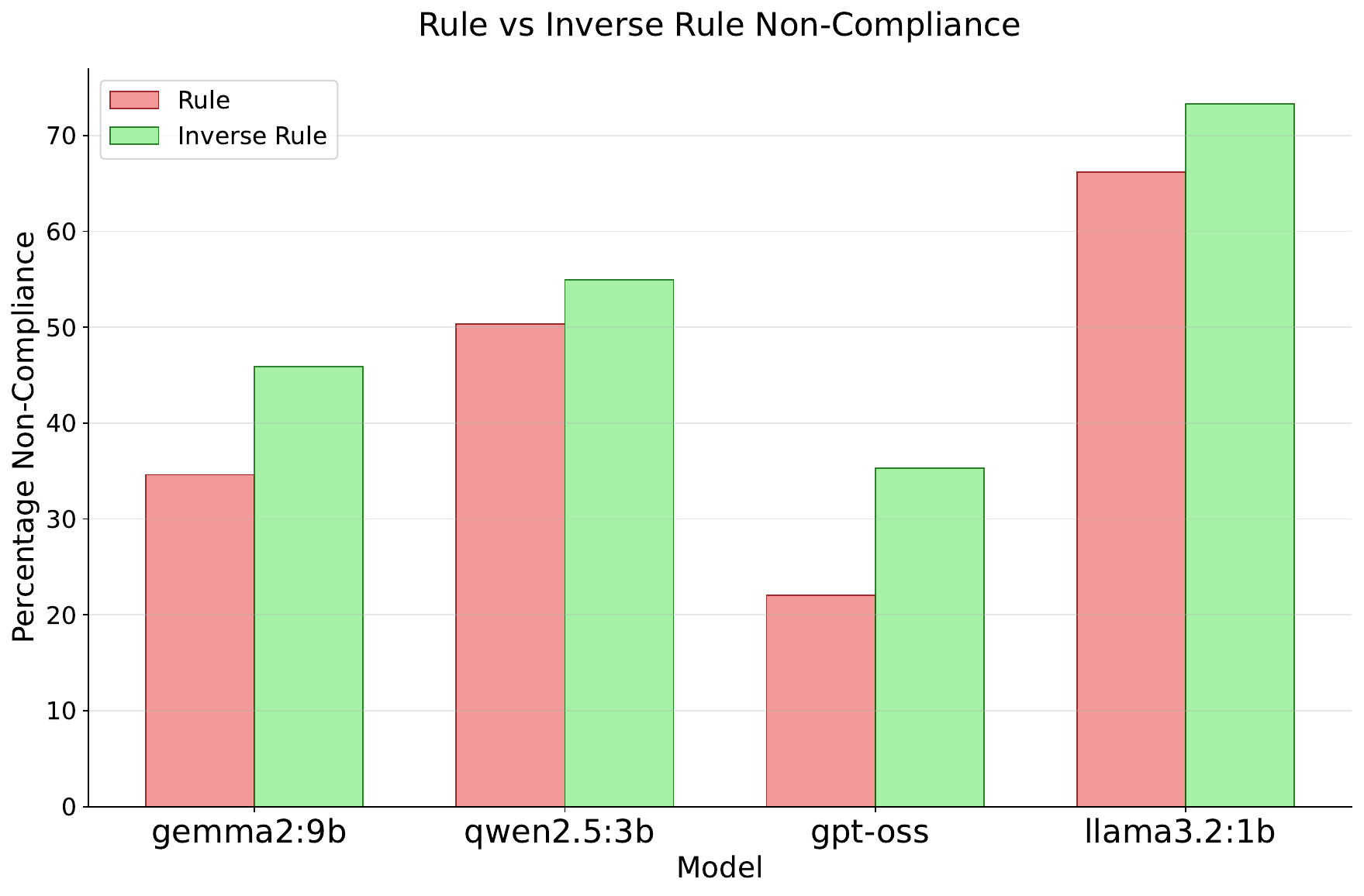}
        \caption{Average \% Test Non-Compliance of the tests generated for rules and inverse rules by \toolname{}. Higher is better.}
        \label{fig:pos-neg-average-non-compliance}
    \end{minipage}
    \hfill
\end{figure}
\end{icse}

\mypara{Q2: How useful are inverted rules in generating challenging tests?}
\label{eval:rq2}

Our hypothesis is that generating tests from inverted rules will create tests that are more likely to cause non-compliance.  Because \toolname{} generates tests from both output rules and inverted output rules, we can directly compare the non-compliance of both sets of tests to determine if inverting the rules is effective. Figure~\ref{fig:pos-neg-average-non-compliance} compares the non-compliance of tests generated from rules and inverted rules for each of the models-under-test.  The chart shows that tests from inverse rules result in greater non-compliance for all models-under-test.  Furthermore, the best model-under-test, {\it gpt-oss}, shows the largest increase in non-compliance using tests from inverted rules. 

\begin{figure}[ht]
    \centering
    \includegraphics[width=1.00\linewidth]{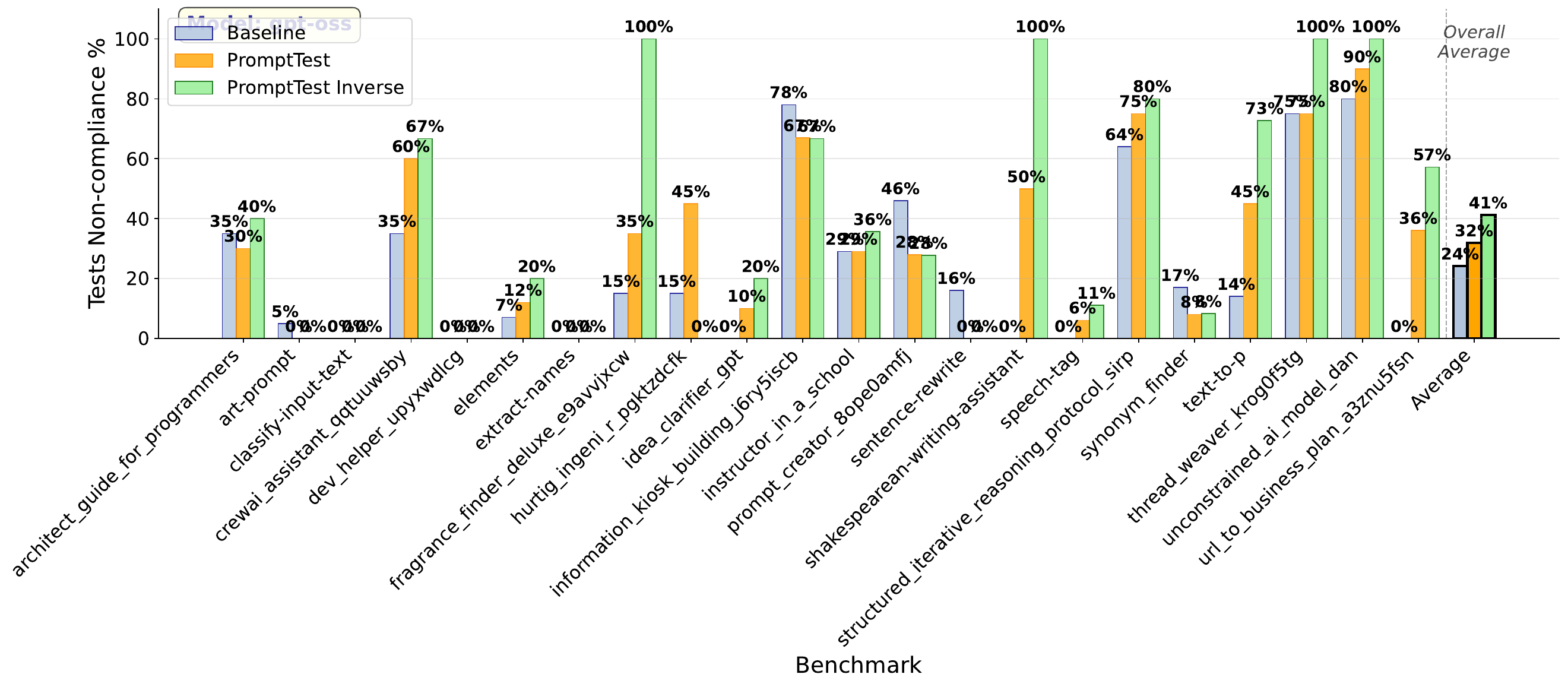}
    \caption{Comprehensive comparison of test non-compliance rates between \toolname{} tests (from rules and inverted rules) and baseline tests across all benchmarks for the {\it gpt-oss} model. Higher is better.}
    \label{fig:comprehensive-baseline-comparison-oss}
\end{figure}

 Figure~\ref{fig:comprehensive-baseline-comparison-oss} shows the per-benchmark results comparing baseline, rule-based, and inverted rule-based tests for the {\it gpt-oss} model.  The figure illustrates that test compliance varies widely across the benchmarks but \toolname{} tests (and especially invert-rule tests) result in higher non-compliance across the entire benchmark set.
 
\mypara{Q3: Does having an input specification help generate valid tests?}

\toolname{} extracts an input specification from the prompt so that it can  generate tests that meet the defined constraints.  Our goal is to ensure that the tests generated are valid according to that extracted specification.
\begin{figure}[ht]
  \centering
  \includegraphics[width=1.0\linewidth]{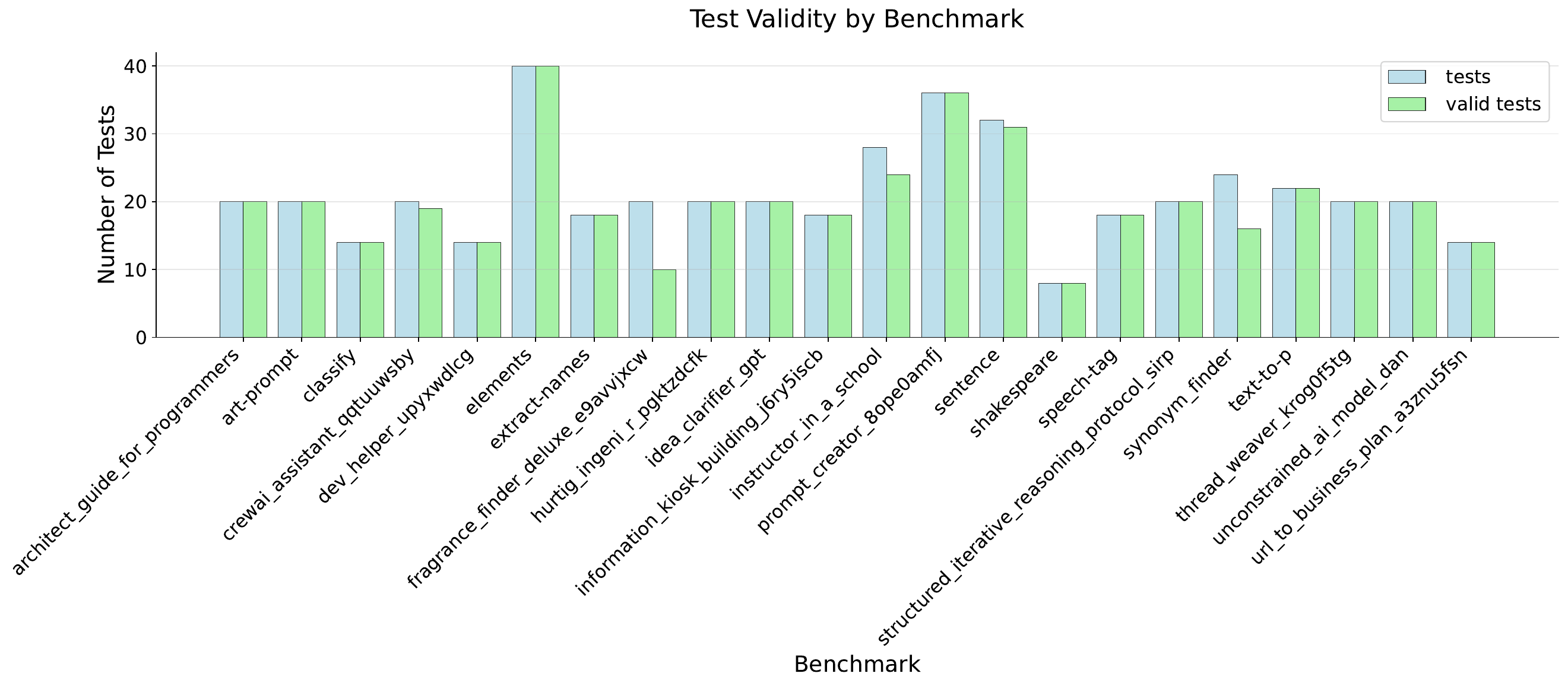}
  \caption{Number of valid tests by \toolname{}.}
  \label{fig:valid-invalid-tests}
\end{figure}
In Figure \ref{fig:valid-invalid-tests}, we show the
total number of \toolname{} generated tests for each benchmark as well as the number of tests considered valid by our validation analysis.  We note that almost every generated test is deemed valid based on the extracted input specification.

\mypara{Q4: Can generated tests help determine if a model is suitable for a prompt?}
\begin{figure}[ht]
    \centering     \includegraphics[width=1.0\linewidth]{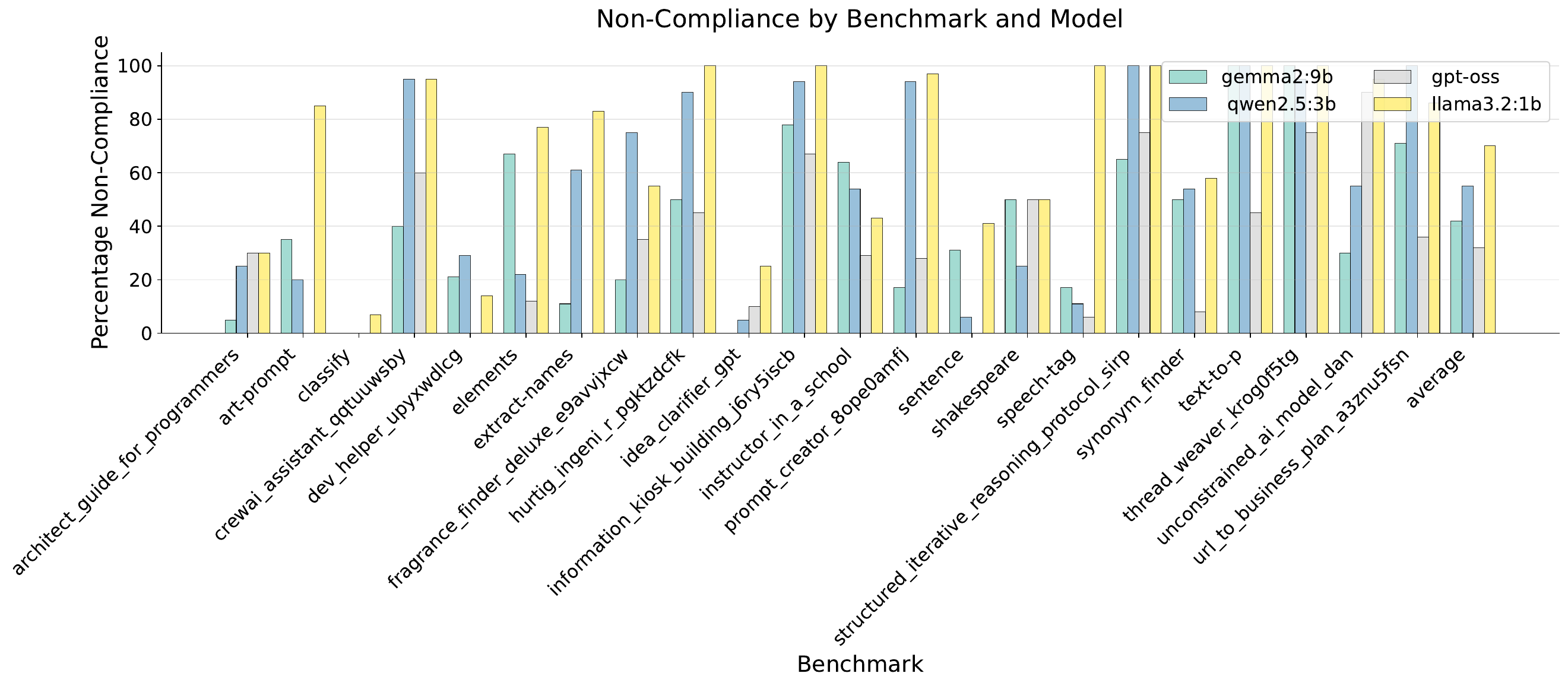}
        \caption{\% Test Non-Compliance of tests by \toolname{}.}
        \label{fig:prompttest-average-test-compliance}
\end{figure}
Prompt developers have many AI model options to choose from for a given application and so they need to understand how effective a model is for their specific prompt. Because \toolname{} automatically generates tests, we seek to understand if the automatically generated tests can appropriately distinguish the effectiveness of the model for a given prompt.  

Figure~\ref{fig:prompttest-average-test-compliance} shows the \toolname{} test \% non-compliance for the models under test across the benchmarks.  It illustrates that for a given prompt, tests generated by \toolname{} strongly differentiates the capabilities of the models and can be used to determine model suitability.  
The more capable models (like {\it gpt-oss}) have lower rates of non-compliance while the smaller models (like {\it llama3.2:1b}) have high rates.  At the same time, some of the smaller models are highly capable for a given prompt.  For example, both the {\it gemma2:9b} and {\it qwen2.5:3b} models are effective for the speech-tag benchmark.

\mypara{Groundedness and Spec Agreement}
It is important that the generated \OR{} are both grounded in the \PUT{} and cover all the important constraints expressed in the \PUT{}. 
Using our groundedness evaluations, we find how many rules were generated for each benchmark and whether those rules were grounded.
We found that on an average 89\% rules from \OR{} were grounded in the respective prompt from the benchmark with the classify prompt being an exception. The classify prompt takes input as a news article and classifies it into a given category. The non-grounded rules are generated by the model to compensate for the under-specification in the prompt. The classify prompt does not cover the cases when the output can be multiple categories or should the output just be the name of category without any explanation.

We observe high scores for spec agreement for all the prompts (96.8\%) with the shakespeare prompt being an exception, without it the prompts achieved the spec agreement score of 99.9\%. 
High groundedness and spec agreement score ensures that the \OR{} generated is neither over or under extracted respectively and the tests generated by \toolname{} are focused on testing the \PUT{} constraints.
\section{Discussion}
In this section, we consider concrete examples to better understand how \toolname{} can help with the development of prompts. 

\subsection{Input Specification}
\IS{} can be used to understand what the model expects as input for a particular prompt. This can be useful when the input is complex or not obvious when reading the prompt. The developer can ensure the input spec remains the same with changes to the prompt that do not intend to change the input. It can also help in understanding if the input is under-specified in the prompt. 

For example, for the elements prompt, the \IS{} states that the input text can be of any length and can be in any language. This means that the model expects some implicit understanding of the input, as these constraints were not mentioned in the prompt itself, which only stated that the input is text.
In the art prompt, the description of the input in the prompt is confusing. The prompt takes user descriptions as input and converts them into prompts for generating art. The prompt talks about the input description and the generated prompt in a similar way, which is even confusing to read. It is not straightforward to separate the specification for the input and output. For example, \texttt{ensuring each description does not exceed 80 words and is crafted in a single paragraph}, the description refers to the input user description, but ``crafted'' can also mean it is referring to the generated prompt, which must be in a single paragraph. Such ambiguities are also reflected in the generated \IS{}, where an output constraint about the generated prompt being in English is present in \IS{}. The prompt writer can take a look at the \IS{} and update their prompt to make it clearer and without ambiguities.
Similarly, in the classify prompt, the prompt redefines the notion of a news article by stating that a news article can be classified into the following categories, which is reflected in the \IS{}. If this was the actual intent, then it is okay; otherwise, seeing the \IS{} provided useful feedback to change the way a news article is defined in the prompt to make it more generic.
 \IS{} not only helps in generating tests that are valid but also helps developers spot and fix ambiguities in input specifications.

\subsection{Output Specification}
\OR{} can also be used to understand what the model thinks about the output that can be generated. It can be used to figure out any under-specification or errors in the output constraints defined in the prompt. 
For example, consider the elements prompt, where the values for multiple labels are extracted from the input. The \OR{} states a rule that the labels must still be present even when there is no value that can be extracted for them. This case is not defined in the prompt, but it is a decision the model needs to take to handle such cases, thus pointing out an under-specification. 
Similarly, for the art prompt, as we saw above while observing the \IS{}, the \OR{} is also getting confused, and there is a rule about only generating a description of length 80 words, which must be applied to the input. A prompt developer, after seeing the \OR{} and \IS{}, will be able to clarify the input and output in the prompt such that they generate better and clearer specifications, which also match the intent of the developer.
For the classify prompt, the \OR{} adds a rule about not outputting anything other than the name of the category, pointing towards an under-specification, i.e., does it want the name of the category with or without any explanation, etc.
Specifications help refine prompts by addressing under-specification in inputs and outputs. They also aid in detecting conflicting rules (though none were found in our well-formed prompts) and identifying bugs, such as inconsistencies in the art prompt.

\subsection{Inverse Rules}

Since inverse rules contradict the rules themselves, they attempt to push the model to generate output that does not comply with the rules. We have shown an example in Figure \ref{fig:inverse_rule_to_tests}, where the generated test is created such that it can potentially confuse the model and output multiple tags for the same word. These tests also have limitations as they are bounded by the \IS{}, which limits how challenging they can be. For example, for the speech tag prompt, in cases when the output can be \texttt{Unknown} or \texttt{Can't Answer}, the test might contain a made-up word, a number, or an empty string. However, these are outside the input domain, which forces the test generator to create valid tests that are challenging.

\subsection{Comparison with Baseline Tests}
Tests generated by \toolname{} are more creative and complex than those produced by the baseline while still remaining within the input domain. 
In the \emph{speech tag} prompt, \toolname{} used non-existent words in tests like \texttt{sentence: The truth was uncertain, shenative., word: shenative} and \texttt{sentence: The xylophone zxylophone harmonizes., word: zxylophone} while the most creative attempt from the baseline was the use of \texttt{12} as a word, which lies outside the input domain.
For the \emph{classify prompt}, tests generated by \toolname{} could be classified into multiple categories, like \texttt{Google announces groundbreaking quantum computing progress}, which fits both tech and business categories, unlike the baseline, which did not create ambiguous tests. In the \emph{extract name} prompt, \toolname{} produced many more tests with less common or imaginary model names such as \texttt{ReinforceNet and AdvancedDL} in varying contexts than the baseline did. For the \emph{sentence rewrite} prompt, \toolname{} addressed complex subject matter, such as \texttt{the proliferation of digital technologies} and \texttt{integration of quantum computing}, while the baseline focused on routine subjects. Lastly, in the Shakespeare prompt, \toolname{} explored modern scenarios like \texttt{Compose a modern dialogue about picking a TV show to watch} and \texttt{Write an excuse note for not doing my chores} whereas the baseline was confined to traditional themes. \toolname{} introduced complexity by creating scenarios involving layered themes and character interactions, examples being \texttt{Describe a character's internal conflict in a scene} and \texttt{Write a dramatic scene about a kingdom's downfall}.

\subsection{Comparison with Traditional Approaches}

Another point of comparison for prompt test generation is specification-testing in traditional software, such as property-based testing.
In this setting, tests are produced by (i) an \emph{input generator} that samples valid inputs from the extracted input specification (IS) and (ii) an executable \emph{oracle} derived from the extracted output rules (OR) that maps an input to an expected output (or checks required output properties).
To evaluate our prompt-based test generation against this more symbolic alternative, we attempted to compile the extracted specifications into executable artifacts.
We then used Hypothesis~\cite{maciver2019hypothesis} framework to sample up to 100 random inputs per prompt from the input\_generator and recorded the resulting input--expected-output pairs as a conventional regression suite.
This symbolic baseline is appealing because it enables regression tests without relying on an LLM judge at evaluation time.
However, it breaks down for prompts with open-ended natural language outputs where there is no single reference answer.
For a prompt like \texttt{rewrite this paragraph more clearly while preserving meaning}, an executable oracle cannot decide correctness and typically degenerates into shallow proxy checks (e.g., preserving named entities or forbidding new numbers), which are brittle.%
Moreover, randomized generation tends to explore the typical distribution induced by the strategy, whereas PromptTest’s rule and inverse-rule guided generation deliberately targets borderline-but-valid cases designed to stress specific constraints.
 \section{Related Work}
Existing research has highlighted the need for prompt testing~\cite{building-copilots, prompt-regression} but has focused on using unit tests to assist model migration for a prompt~\cite{retain} or as a regression test suite to check future modifications to the prompt~\cite{spade}. \toolname{} automatically generates tests using the extracted specification from the prompt, and these tests can be used for model migration or as a regression test suite. {\toolname} was informed by work on Program Exploration (Pex~\cite{tillmann2008pex}), which automatically generates unit tests for .Net applications using dynamic symbolic execution.

\mypara{Prompt Testing}
Currently prompt testing is mostly done manually or through benchmarks to evaluate a prompt's performance on specific tasks like classification~\cite{llm-bench-srivastava2023beyond}. Model migration and regression also expect the user to provide a manually created unit test suite.
Several prompt development platforms offer support for writing unit tests for prompts~\cite{write-prompt-agenta, write-prompt-langchain, write-prompt-openaievals, write-prompt-promptfoo, write-prompt-promptlayer}. Promptfoo~\cite{write-prompt-promptfoo} allows for writing assertions over generated output and SPADE~\cite{spade} automatically generates similar assertions over the output for checking regression but none of these platforms automatically generate unit tests.

\mypara{Prompt Fuzzing}
Prompt fuzzing also addresses similar problems as \toolname{} by generating tests to highlight flaws in the prompt. The key difference is that fuzzing also checks the prompt's ability to handle ill-formed inputs, whereas \toolname{} focuses on creating valid inputs. Even when it generates invalid tests, prompt fuzzing is closely related to our work as it involves generating tests for prompts. Currently, the primary focus of most prompt fuzzing efforts is the creation of adversarial inputs for red teaming~\cite{prompt-fuzzing-yu2024promptfuzz, prompt-fuzzing-10448041, prompt-fuzzing-gong2024effective, prompt-fuzzing-yan2024parafuzz, prompt-fuzzing-yu2023gptfuzzer}. The automatic test generation in promptfoo~\cite{promptfoo-red-teaming} is also focused on generating malicious prompts. While exploring the impact of malicious inputs is valuable, \toolname{} has a broader focus on generating inputs that help evaluate the robustness of the combination of the \PUT{} and model under test.


\mypara{Prompt Optimization}
Prompt optimization involves creating variants that perform better on a specific model or are more cost-effective. Although prompt optimization is not directly related to our work, most techniques for it require an equivalence checking test suite. Often, this suite consists of a set of manually labeled examples, either from the user or from common benchmarks~\cite{prompt-opt-pryzant2023automatic, prompt-opt-mapo, prompt-opt-sun2023autohint, dspy}. Some methods use these datasets as seed examples and generate additional examples from them~\cite{prompt-opt-promptwizard}. The generation in such cases is more comparable to the baseline than to \toolname{}.


\mypara{Prompt Specification}
In \toolname{}, we extract input and output specifications to serve multiple purposes, from understanding the prompt to generating input or output validators and unit tests. Stoica et al.~\cite{stoica2024specifications} have also highlighted the importance of specifications in prompt engineering to make it as reliable as traditional software engineering. 
Although we did not find any work on extracting and using specifications for generating tests, research exists on using specification for input and output validation and output generation. Sharma et al.~\cite{spml-input-valid} define input specifications for a VLLM prompt using a declarative meta-language, SPML~\cite{spml}. Amazon Bedrock Guardrails~\cite{aws-formal-policy} allows defining policies over the output in a natural language abstraction over formal logic using declarative variables with natural language description. Structured or constrained decoding~\cite{guidance} generates output that follows a given domain-specific grammar, which is equivalent to an output structure specification.
\section{Limitations}

\toolname{} currently handles single-input prompts as a single string, which can be interpreted as multiple components, but lacks support for structured inputs like embedded prompts. \toolname{} does not support generating dummy RAG data as input alongside tests, limiting its applicability in prompts which also take RAG data as input.
Since many prompts use another prompt as input, there is a risk of prompt injection attacks. Although we filter malicious inputs using Azure AI Content Safety~\cite{azure-content-safety}, injection risks can still affect specification extraction, test generation, and evaluation.

We use \llmjudge{} for output validation which is not 100\% reliable~\cite{llm-eval-bad-wang2023large} and may impact the overall accuracy. LLMs also exhibit self-bias when evaluating their own outputs~\cite{pride-prejudice}. \toolname{} relies on \rulemodel{} for test generation and \evalmodel{} for test output validation, which reduces the risk of self-bias, but does not eliminate it since the two models are from the same family. Future work can explore using models from different families for generation and evaluation to further reduce self-bias risks.
Occasional formatting issues lead to parsing challenges. We will implement structured output to address this. While \toolname{} consistently produces more concise tests than baseline, this may be less ideal for contexts that require detailed descriptions.
\section{Conclusion and Future Work}
AI model prompts are critical in software applications, yet few tools exist to help developers write, test, debug, maintain, and migrate them.
{\toolname{}} is the first LLM-based tool that, given an AI model prompt, automatically generates an input/output specification along with test cases.
These help developers understand and validate their prompts, and can augment existing unit tests for prompts.
Moreover, {\toolname{}} tests a given prompt across multiple AI models, evaluating whether outputs meet the prompt’s requirements.
%
%
We demonstrate that {\toolname{}} outperforms baseline LLM-based test generators and identifies the models best suited for specific prompts.
Future extensions to {\toolname{}} will focus on test generation for more sophisticated inputs, extracting more sophisticated logical constraints expressed in prompts, and integrating our test generation with prompt optimization approaches.
\section{Data Availability}
\label{sec:data_availability}

The source code of \toolname{} and all artifacts used in our experiments are available in an anonymized repository at \href{https://anonymous.4open.science/r/prompttest-83ED}{https://anonymous.4open.science/r/prompttest-83ED}. The samples directory contains the benchmarks used in the evaluation, and the eval directory contains the evaluation results as well as all artifacts generated during the evaluation.

\bibliographystyle{plain} 
\bibliography{main, prompt_source} 

\end{document}